# The Weak Reality that Makes Quantum Phenomena more Natural: Novel Insights and Experiments

Yakir Aharonov [1 2 3 4], Eliahu Cohen [5 4], Mordecai Waegell [1], and Avshalom C. Elitzur [4 1]*

[1] Institute for Quantum Studies, Chapman University, Orange, CA 92866, USA

[2] Schmid College of Science and Technology, Chapman University, Orange, CA, 92866, USA

[3] School of Physics and Astronomy, Tel Aviv University, Tel Aviv 6997801, Israel

[4] Iyar, The Israeli Institute for Advanced Research, POB 651, Zichron Ya'akov, 3095303, Israel

[5] Faculty of Engineering and the Institute of Nanotechnology and Advanced Materials, Bar Ilan University, Ramat Gan 5290002, Israel

*Corresponding authors: avshalom@iyar.org.il

*While quantum reality can be probed through measurements, the Two-State-Vector formalism (TSVF) reveals a subtler reality prevailing between measurements. Under special pre- and post-selections, odd physical values emerge. This unusual picture calls for a deeper study. Instead of the common, wave-based picture of quantum mechanics, we suggest a new, particle-based perspective: Each particle possesses a definite location throughout its evolution, while some of its physical variables (characterized by deterministic operators, some of which obey nonlocal equations of motion) are carried by "mirage particles" accounting for its unique behavior. Within the time-interval between pre- and post-selection, the particle gives rise to a horde of such mirage particles, of which some can be negative. What appears to be "no-particle," known to give rise to Interaction-Free Measurement, is in fact a self-canceling pair of positive and negative mirage particles, which can be momentarily split and cancel out again. Feasible experiments can give empirical evidence for these fleeting phenomena. In this respect, the Heisenberg ontology is shown to be conceptually advantageous compared to the Schrödinger picture. We review several recent advances, discuss their foundational significance and point out possible directions for future research.*

## 1. Introduction

For many years, the Two-State-Vector Formalism (TSVF) [1-3] has been unearthing more and more hidden aspects of quantum reality never conceived before. The basic premise is simple: Quantum theory, like classical physics, is time-symmetric, save for the "wavefunction collapse" introduced by measurement. This gives the notion of quantum measurement a profound twist. The measurement's effect goes not only forward in time but backwards as well. Consequently the particle's physical properties between two measurements are affected by both past (pre-selection) and future (post-selection) effects. The resulting picture is fully consistent with standard quantum theory yet reveals hitherto unnoticed aspects of the process, namely "weak values" [4-7]. The latter constitute a "weak reality" which offers a deeper understanding of quantum reality and how it is related to the classical one [8-10].

The underlying mathematics is simple and intuitive. To determine some physical property A of the system at time $t$, we evolve the initial state $|\psi(t_i)\rangle$, prepared at time $t_i < t$, from past to future, and then evolve the final state of the system $|\phi(t_f)\rangle$, determined at a later time $t_f > t$, from future to past. We then combine at each moment $t$ the two evolutions using the "two-state" $\langle\phi(t)|\ |\psi(t)\rangle$ to infer the weak value of any operator A defined as:

$$\langle A \rangle_w(t) = \frac{\langle \phi(t) | A | \psi(t) \rangle}{\langle \phi(t) | \psi(t) \rangle}. \tag{1}$$

Such values can be extraordinary – very large, very small or even complex, lying outside the spectrum of the measured operator *A* [9-14]. These weak values manifest themselves as effective interaction terms between the pre- and post-selected system (and any other system coupled to it weakly enough) [15]. Yet they prevail *between* rather than *upon* quantum measurements, thereby being inaccessible to direct inference via the standard measurement techniques. Several methods, described below, were invented to bypass this difficulty, and eventually vindicated a plethora of surprising predictions, see e.g. [4-20]. TSVF is therefore much more than just an interpretation of QM. On the one hand it is fully consistent with the conventional, one-vector formalism, hence all its predictions are obliged by the latter as well. Yet this affirmation by the conventional formalism always comes with hindsight. In other words, none of the TSVF's intriguing predictions have ever been proposed by the standard approach! This computational efficiency lends support to TSVF's ontological soundness as well.

Gradually, a broad and self-consistent landscape began emerging from the formalism and its offshoots. Weak values, it turns out, *underlie* the ordinary quantum values [10,21,22], offering a novel yet very natural explanation to quantum oddities considered so far axiomatic or even banned from realistic inquiry by Copenhagen-like interpretations.

A deeper understanding of the dynamics of these weak values is offered by "mirage particles," momentary particles springing from the initial particle during the above "between-measurements" interval. This concept was already alluded in earlier works of ours [14,19-22] and colleagues [23,24]. Among these mirage particles there are some whose very presence has a minus sign, implying that, upon a weak enough interaction, their properties, including mass and charge, reverse their sign [15,25,26]. We also refer to mirage particles having negative weak values as "nega-particles" [14]. The formalism shows how positive and negative mirage particles can cancel one another into an apparent "nothing" [21,22], somewhat similarly to particle-antiparticle annihilation but with no energy output, and with the possibility of parting again out of the vacuum. Feasible experiments, awaiting laboratory realization, have already been proposed for demonstrating these predictions, some of which are described below.

The consequences for quantum theory are far-reaching. The particle's hypothesized multiplication, disappearance and reappearance prior to measurement [19,20] offer an intuitive account of the wavefunction's oddities like "collapse," nonlocality and temporal anomalies. The present article sketches this evolving formalism of quantum mechanics and points out new directions for future research.

The paper's outline is as follows. Section 2 uses the simplest quantum mechanical type of measurement for illustrating TSVF's approach. In Sec. 3 we describe an interesting prediction of TSVF related to an extensively studied nested Mach-Zehnder Interferometer (MZI) setup, where mirage and nega-mirage particles are involved, and the challenges it poses to theory and experiment. Sec. 4 presents two customary methods of validation, namely weak measurements as well as projective (strong) ones. Sec. 5 describes an equivalent experiment that illuminates additional aspects of mirage particles' dynamics. In Sec. 6 we further discuss the physical meaning of negative weak values. In Sec. 7 we present a very preliminary outline for generalizing this formalism, and point out further avenues for research. In Sec. 8 we show how these predictions are best understood using the Heisenberg particle-based rather than the Schrödinger wave approach.

**2. How "Void" are the Wavefunction's Non-Observed Parts?**

For an intuitive introduction to TSVF consider the simplest quantum-mechanical measurement setup. One photon hits a beam-splitter, its wavefunction splitting into transmitted and reflected halves, and finally detected as a single photon by one of two equidistant detectors (Fig. 1).

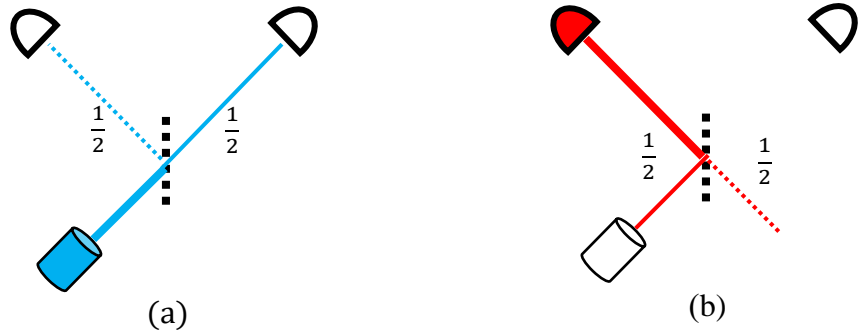

**Figure 1.** A particle split by a beam-splitter is predicted to go through one out of two possible paths and eventually be detected in one, the other becoming "void" (a). Similarly for the time-reversed retrodiction (b): The wavefunction splits again towards the past, one half leading to an obviously void "origin."

There are two "void" branches to this process. One is familiar, namely i) the path leading to the detector that eventually did not click (1a). A subtler one branches from the backwards path, returning from the clicking detector to the beam-splitter: ii) One half returns to the source but the other goes to the opposite direction from which the photon could have never come (1b).

These two void parts of the particle's evolution epitomize quantum mechanics' two major contrasts with classical physics, namely i) indeterminism and ii) the time-asymmetry inflicted by measurement. They seem to present mere mathematical curiosities with no physical content. TSVF, however, can extract from them a surprising physics: A combination of such future and past void branches within one evolution gives rise to a temporary particle in a location where it seems to have never gone. Other unusual phenomena then follow, described in the following sections.

**3. Can a Particle be where it Never Went?**

Consider [16,17] an MZI within which a smaller one is nested (Fig. 2). The first beam-splitter $BS_1$ splits the beam into 1/3-2/3, and the last, $BS_4$, *vice versa* into 2/3-1/3. On the right, 2/3 arm E, lies a smaller, standard MZI with two 50% BSs.

Let a photon go through the setting. This preparation gives rise to the initial state:

$$|\psi\rangle = \frac{1}{\sqrt{3}}(|A\rangle + |B\rangle + |C\rangle). \qquad (2)$$

If the photon takes the right path E and enters the smaller MZI, then, by constructive interference, it must exit towards detector $D_1$ and never make it through F to the final $BS_4$ and the last two detectors $D_2$ and $D_3$.

Select, then, the cases where $D_1$ did not click. This entire part of the wavefunction now becomes "void" in the sense of the previous section: The photon seems to have never taken this E arm, but to have rather taken C.

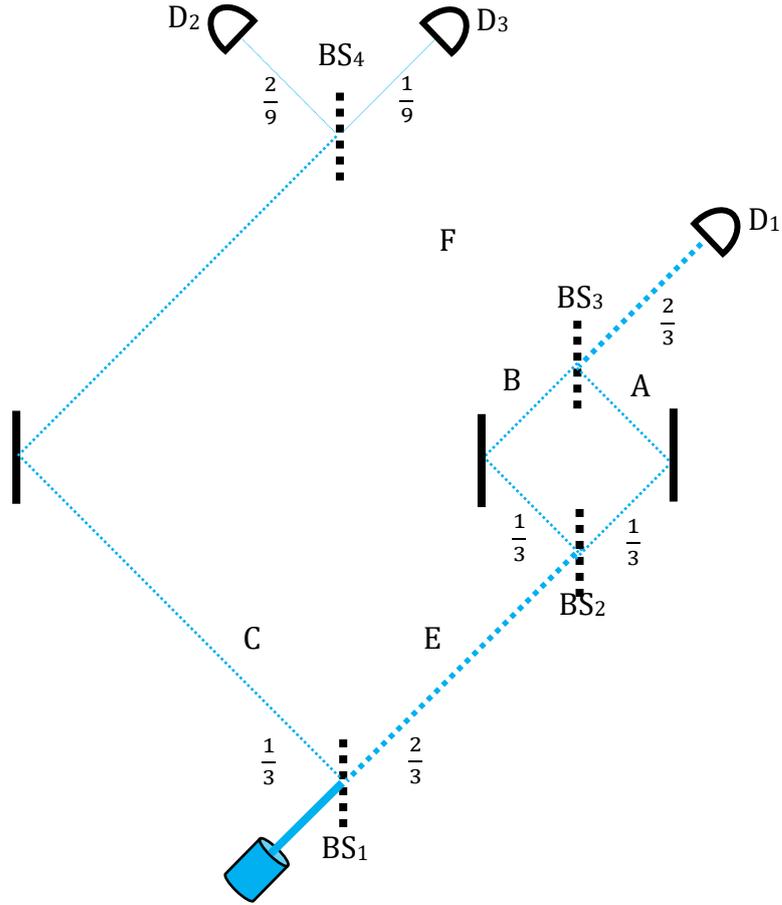

**Figure 2.** Vaidman's nested MZI [16,17]. From BS$_1$ the path goes to a smaller MZI between BS$_2$ and BS$_3$. The path emerging from the nested MZI in case of constructive interference goes to detector D$_1$, whose non-clicking cancels the entire right-hand path, implying that the photon never passed BS$_1$ but was rather reflected to the left towards BS$_4$ and detectors D$_2$ and D$_3$.

Next proceed to select the remaining 1/9 of cases where D$_3$ has clicked. This amounts to post-selection of

$$|\phi\rangle = \frac{1}{\sqrt{3}}(|A\rangle - |B\rangle + |C\rangle). \tag{3}$$

Then, again by interference, this backward state-vector "leaves" the nested MZI through another "exit," say towards a wall, which, of course, could have never been the photon's source.

However, on this segment of its way back to the past (see Fig. 3), this void branch is going over the earlier void part, that of the forward-moving wavefunction which came from the source through arm F into the nested MZI:

$$|F\rangle \to \frac{1}{\sqrt{2}}(|A\rangle - |B\rangle). \tag{4}$$

.

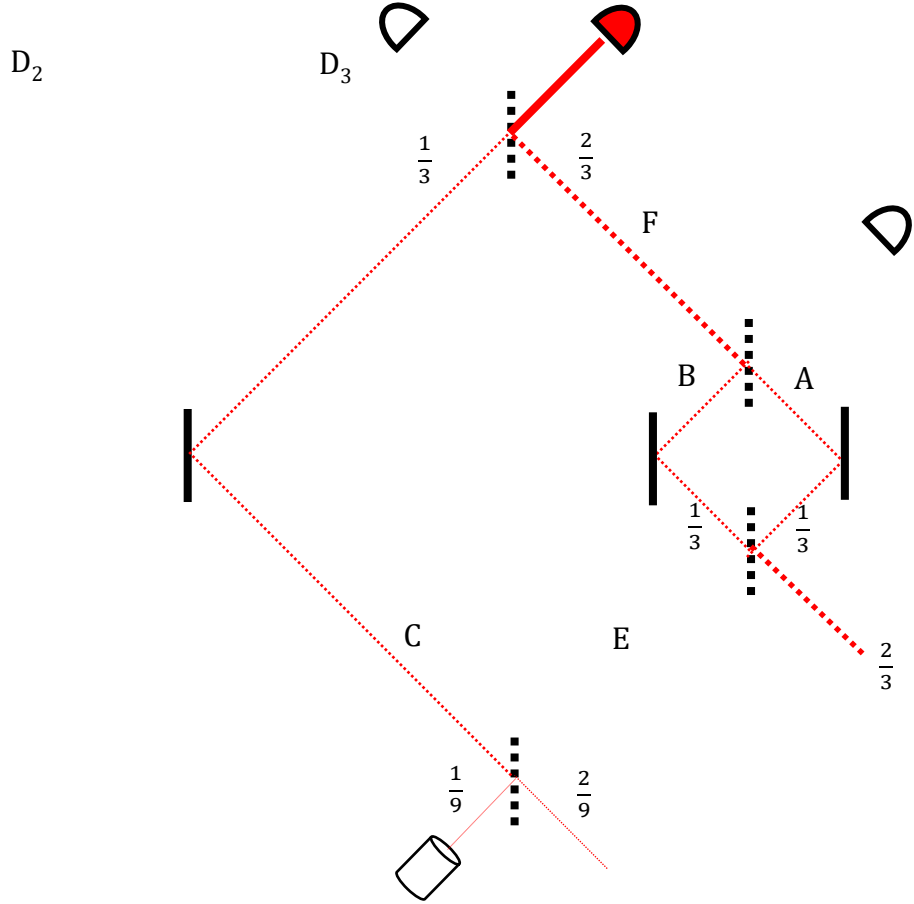

**Figure 3.** The backward evolution from the actual detection. Again a void branch goes to the nested MZI and exits towards an obviously void source.

Combining (2) and (3), then, gives a surprising result in the form of the two-state:

$$\langle\phi|\ |\psi\rangle = \frac{1}{3}\big((\langle A|-\langle B|+\langle C|)(|A\rangle+|B\rangle+|C\rangle\big), \tag{5}$$

indicated by the minus sign assigned to the particle in the B arm (while the particle must be found in either A or C with certainty if we look for it there).

The corresponding weak values, as defined in (1), are

$$\langle\Pi_B\rangle_w = -\langle\Pi_A\rangle_w = -\langle\Pi_C\rangle_w = -1, \tag{6}$$

$$\langle\Pi_E\rangle_w = \langle\Pi_F\rangle_w = \langle\Pi_A\rangle_w + \langle\Pi_B\rangle_w = 1-1 = 0, \tag{7}$$

where for all $i$, $\Pi_i$ is defined as a projection operator onto arm $i$ of the interferometer (amounting to the question: if we look for the photon in arm $i$, will we find it there?).

In other words, in the middle segment of the right-hand path, where the two void histories overlap, an additional *(*detectable*)* particle appears (Fig. 4). The appearance is short-lived: Only along the right path A of the nested MZI, with no entry neither an exit into and from it!

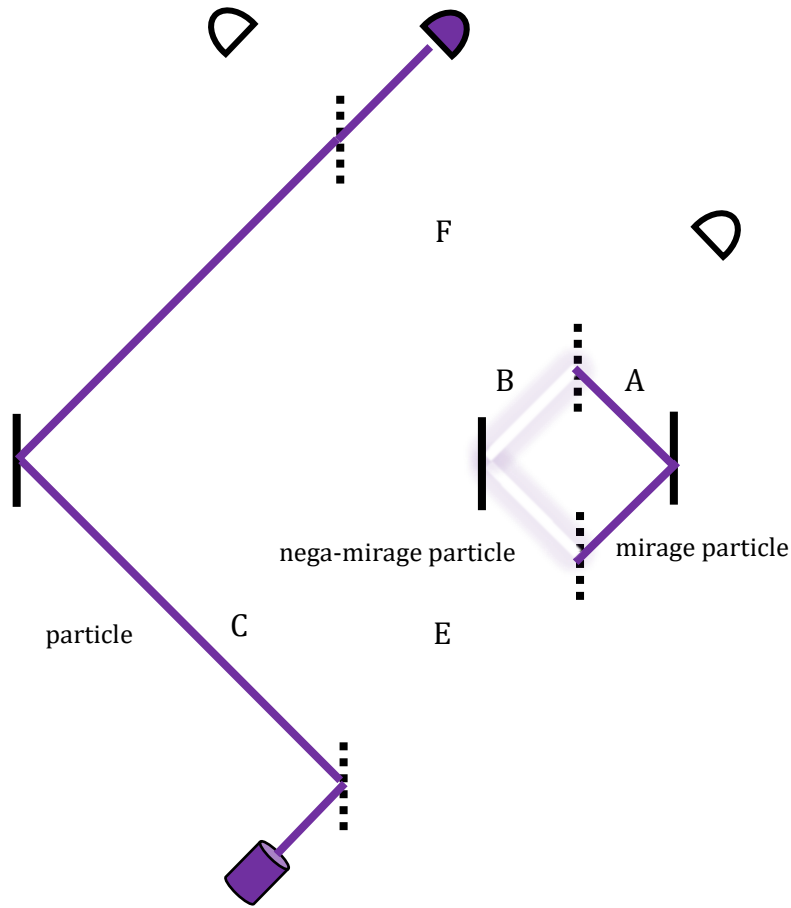

**Figure 4.** The purple trajectory is the overlap of the blue and red lines of the earlier forward and backward state-vectors. A momentary additional particle appears on the right path's middle segment.

## 4. In Search of Validation: Weak and Strong Measurements

This is certainly a striking derivation. Once $D_1$ has failed to click, one would regard this part of the wavefunction completely ruled out, "collapsed" into nothingness. Lo and behold, when $D_3$ later clicks, then, within the middle segment of this never-traversed trajectory, the particle is revived!

Striking indeed, yet apparently banned from validation by the fact that this is a retrodiction, holding for the *past*, prior to the final click. In other words, the derivation holds only if we have refrained from measuring the particle's whereabouts within the MZI. Is this derivation, then, doomed to remain inaccessible to empirical proof?

One bypass is offered by weak measurement [4]: Let the coupling between particle and detector be very weak, thereby highly plagued by quantum uncertainty. Upon sufficiently many trials, the averaged result gives the weak value with arbitrarily high precision yet with no visible disturbance. Several such experiments have already been carried out, and all TSVF predictions have indeed been verified by this method.

What would weak measurements reveal for the present case? Let us (gedankenly, ignoring technical issues) make the two solid mirrors of the nested MZI small enough and movable, such that they can react with the slightest recoil to the photon's taking the right or left path. Post-select for all cases where $D_3$ has clicked, apparently implying that the nested MZI has never been traversed. Because the mirrors' momenta are subject to quantum uncertainty, repeat the experiment sufficiently many times to overcome the noise. The predicted result offers the first affirmation to the TSVF prediction, moreover a double one: The right-hand mirror indicates a recoil upon the photon's overall hits, while the left hand undergoes a negative recoil, namely a "pull" rather than a "push"!

Skeptics, however, have objected to weak measurements as a means for revealing true quantum properties of the system [27-29] (see however the reply in [30]), often explaining away their outcomes as noise inflicted by the measuring device's uncertainty. Even stronger objections have been raised against Vaidman *et al.*'s version of weak measurements in the present case [17] because it has employed classically vibrating mirrors and classical beams (rather than single photons), see e.g. the discussions in [31-34].

"Extraordinary claims require extraordinary evidence" [35]. While we find the objections to weak measurements ill-reasoned, we want to face the challenge head-on. A particle's alleged fleeting appearance, in the middle of a path it seems never to have entered nor exited, is extraordinary enough to merit a more unequivocal validation. Such a validation will, in turn, add credibility to weak measurements as well.

Fortunately, such a method has been introduced, and moreover for a TSVF prediction analogous to the nested MZI. This is a standard, projective measurement, hence immune to all objections against weak measurement. First, here is a brief account of the method's development.

Okamoto and Takeuchi [36], following an earlier suggestion of Aharonov and Vaidman [37], have realized, using a novel photonic quantum router, a photon that acts like a "shutter" that reflects a probe photon "hitting" it. They took the TSVF analysis of a photon superposed over three locations, where, upon the appropriate post-selection, it is predicted to act as a shutter, with certainty, in two of them at the same time. A probe photon, also superposed, directed towards the superposed shutter, has become entangled with it, as if being reflected from both locations. Here, then, is a TSVF prediction verified with a *standard* quantum measurement. Elitzur *et al.* [20] took this technique one step further for testing, with a finer temporal resolution, another intriguing TSVF retrodiction [19]. The experiment involves a similar three-boxes setting, within which the particle is retrodicted to disappear and reappear at different instances across distant boxes. A probe photon, this time superposed in both space and time, interacts with the three boxes at the times the shutter photon is supposed to be present and absent. The two photons become correlated only if the shutter reflects the probe photon when the former is expected to be present, and lets the probe photon pass through its box when absent. As these instances of the shutter's presence and absence occur one after another in the same boxes, it seems to have disappeared and reappeared time and again.

This method of validation can be applied to the present setting, namely the nested MZI. A feasible optical setup was given in [20], hence the following discussion is on the pure gedanken level (Fig. 5). Let the photon going through the entire nested MZI device be a shutter photon. Let a probe photon be split in both space and time such that it interrogates the whereabouts of the shutter within the device over time. For this purpose the four split branches of the probe photon go, one by one:

*i*) at $t_1$ to a mirror placed just behind the trajectory E leading to the nested MZI;
*ii*) at $t_2$ to the nested MZI's right-hand path A where the mirage photon is expected to be;
*iii*) at $t_2$ to the large MZI's left path C where the photon is simultaneously expected to be;
*iv*) at $t_3$ to a mirror placed behind the exit trajectory F from the nested MZI towards $BS_4$.

We expect the probe photon to hit the first mirror without being disturbed by the any shutter photon on its way to the nested MZI; then to be reflected by both the mirage photon within the nested MZI and the shutter photon on the large MZI's left path; and then again to be reflected by the second mirror, indicating that no shutter photon has left the nested MZI along F. The resulting shutter-probe entanglement is Bell-like: One can either check correlations between their paths, or between their interference patterns [20]. Upon a successful post-selection, the latter option would indicate a constructive interference at $D_5$ of all the probe's wavepackets returning from E, A, C and F with their original amplitudes $\alpha_i$, that is

$$|\psi\rangle_p = \left(\alpha_1 |E_T(t_1)\rangle + \alpha_2 |A_R(t_2)\rangle + \alpha_3 |C_R(t_2)\rangle + \alpha_4 |F_T(t_3)\rangle\right), \tag{8}$$

where *T* denotes transmittance through a void part and *R* reflection from a particle.

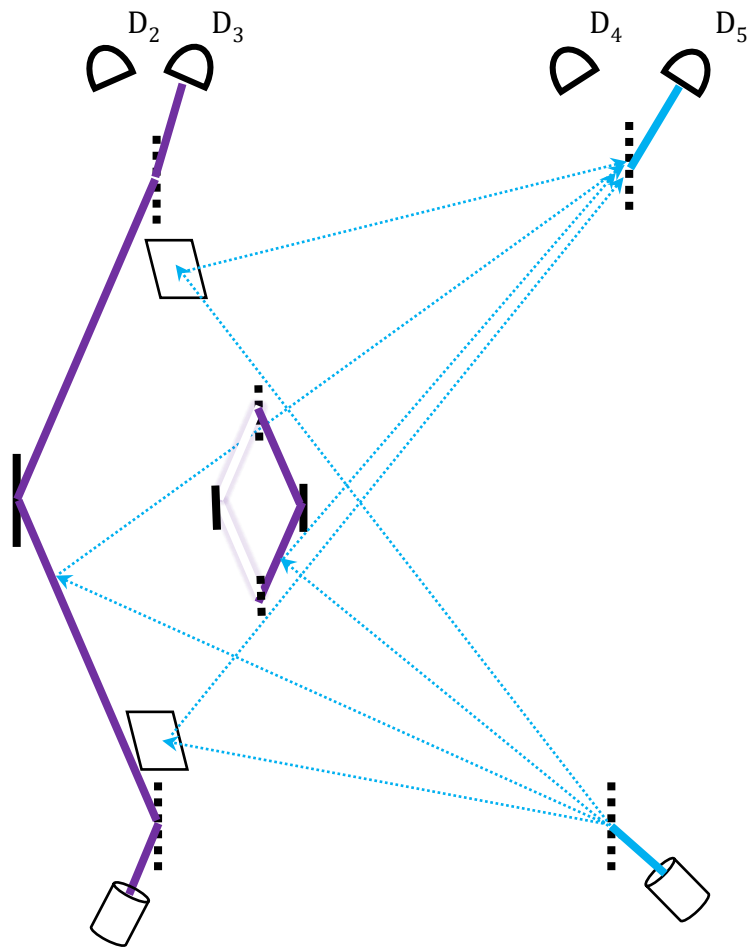

**Figure 5.** A probe photon (drawn in blue), superposed in both space and time, interacts via quantum routers with the photon traversing the nested MZI in three moments at four places where the shutter photon is expected to be either present (the probe being reflected by the shutter) or absent (the probe being reflected by a mirror). Correlation between probe and shutter detectors, $D_2$-$D_5$ and $D_3$-$D_4$, respectively, emerges only if the probe photon is reflected by the mirage shutter photons where they are expected to pass, and by the mirrors where no photon is expected.

Let us conclude with the foundational significance of this experiment. Whereas the photon seems to have never taken the MZI's right path, TSVF reveals a much deeper account. *This path has been taken by a pair of mirage and nega-mirage photons, which through mutual cancellation gave the appearance of no particle*. Vaidman's setting [16,17], as well as [19,20], thus enable a momentary *resolution* of this apparent "nothing" into its two subtle components, followed again by self-cancellation.

Of special interest is the right path's F segment, leading from the nested MZI to final $BS_4$ and detectors $D_2$ and $D_3$. The photon is not expected to pass there by the basic laws of optics, and indeed the probe photon's 4th part is expected not to find it there. Yet this segment should remain *open* – any obstruction along it would make the experiment fail [17]. Why? The TSVF answer is straightforward: It should remain open for the *future* effect of the post-selection at $D_3$. This account, while demanding a great conceptual sacrifice, is the most intuitive for us.

**5. The "Spooky Particle" Experiment**

The nested MZI is a variant of an earlier version, formulated in the form of a particle in three boxes [8]. That experiment (see Sec. 4 above) has been described in great detail in [20] (see also the popular version [38], which gave it its name), so suffice it here to mention only its most salient features. A particle is initially superposed over three boxes, of which A and B are close enough to allow it to move between them, while C is arbitrarily far away. Here the dynamics is cyclic. Under appropriate pre- and post-selections, the particle is expected to reside:

*i)* in A and C at $t_1$;

*ii)* only in C at $t_2$;

*iii)* in B and C at $t_3$;

*iv)* and then again in A and C at $t_4$, *etc.*

In other words, the particle totally vanishes from A and B, to reside only in C (where it could never tunnel), then returns to B, then tunnels back to A and then all over again. Here too, the disappearance-reappearance cycle is due to the nega-mirage particle being first in B, then joining the positive mirage particle in A to make it disappear, then parting from it to make it reappear.

There is an additional intriguing feature to this setting. Consider the time $t_1$ when the two mirage particles coexist in A and C (as already implied by the Okamoto-Takeuchi experiment [36]). The nega-photon is now in B. At this instant, the following retrodiction holds: *Had we joined B with C rather than with A, the particle would have vanished from C and "collapsed" into A.*

This is a unique situation. So far, the two possible measurement outcomes "click" and "no click" were random. Here, however, it is possible in retrospect to point out their cause. Can this insight be generalized to all measurements? To this question we have at present only a few hints, proposed in the following sections.

Significantly, this derivation of the click/no-click "cause" is only *retrospective*. This holds for all TSVF predictions, such as the above disappearance-reappearance. The reason is clear: Having access to phenomena associated with weak values in real time would entail the bluntest causality violations.

**6. On Negative Weak Values: Can a Mirror be "Pushed" Inwards?**

In the weak measurement version of the above nested MZI experiment, we have encountered a curious negative recoil of the MZI mirror that the nega-photon is expected to hit. A similar derivation has been presented in earlier works [25,26,39]. This effect of weak reality adds another piece to our emerging picture, to which we have earlier referred to in a paper titled "1-1=Counterfactual" [22]. Quantum non-events, known for their curious causal efficacy, can be better understood as a sum of positive and negative weak values.

Indeed, as discussed earlier in [13], negative weak values are quite abundant: whenever the weak value of some projection operator exceeds unity, there exists at least one other projection operator with a negative weak value. Only for a measure zero of post-selected states, no negative weak values are expected.

Negative recoil is not the only inverse effect of the nega-particle. For example, when photons are absorbed by excited nega-atoms in the process of stimulated emission, the nega-atoms become ground without a subsequent emission. Alternatively, spontaneously emitted radiation from excited atoms, some of which having negative weak values, can have an extraordinary spatial distribution, indicating interference between photons emitted from nega-atoms and positive ones. Such phenomena are analyzed in detail in [14].

The effects of nega-particles go even further: Weak values determine the effective potential whenever a weak coupling to a pre- and post-selected system is created [15]. Consequently, a negative value of some operator *A* implies that when we couple weakly to *A*, the sign of the interaction term is flipped. In the current paper this physical understanding is attached to all nega-particles, but in [13] an interesting alternative was proposed: Under two plausible consistency conditions (which create the connection to the standard formalism of quantum mechanics), all strange weak values (not only the negative ones) can be interpreted as complex conditional probabilities corresponding to counterfactual scenarios.

In our perspective, then, all weak values are real in the sense that they represent physical, measurable quantities (which were in fact measured in numerous experiments with various systems and multiple methods). However, some of the mirage and nega-mirage particles have a fleeting, transient existence, like in Vaidman's nested Mach-Zehnder experiment or in the authors' "Case of the disappearing particle". Our next step is to explore some generalizations of the proposed dynamics.

## 7. Generalizing: Interaction-Free and Positive Measurements as Sums of Weak Values

The plethora of weak values revealed by TSVF, of which we presented here only a few, can now be viewed in a broader context. These values, we submit, constitute a coherent weak reality that *underlies* quantum reality, thereby offering novel insights into the latter's riddles. For a given pre- and post-selected ensemble, the weak value of every projector in the system's Hilbert space is defined, and this produces a noncontextual value assignment to all observables of the system, which necessarily includes complex and negative values for some projectors.

Quantum reality, on the other hand, as known from measurements, is mostly discrete: elementary particles are indivisible. Yet the formalism describes a continuum underlying reality: A wave propagates in a deterministic fashion, only to give rise to a discrete and indeterministic outcome, namely a particle in an unpredictable location, upon measurement. Weak reality now makes the dual picture much richer. Prior to measurement, mirage copies of the particle, some of which may be too large, too small or even complex, can momentarily appear between pre- and post-selections, and their varying interactions with one another determine the particle's familiar quantum value upon measurement.

How does this transition occur? Let us begin with the case where measurement indicates that the particle is *not* at that location. It is one of the wonders of quantum mechanics that, unlike classical physics, this apparent non-event is not devoid of causal efficacy: The interaction that has not occurred exerts nonlocal effects on the entire wavefunction just as if it has. This is Interaction-Free Measurement (IFM) [40], where even a detector's *non-click* destroys the interference. Within the TSVF, however, this is very natural. Consider again the nested MZI (Fig. 4): What appears to be no particle hitting detector $D_1$ and no detection, turns out to be a self-cancelling pair of mirage particles, of which one is a nega-particle. The nested MZI, then, enables a momentary resolution of this "nothing." Suppose for example that detector $D_1$ is a movable mirror, which, by not recoiling, indicates that the particle did not go that way. According to the TSVF, the mirror's non-recoil is simply the sum of positive and negative recoils. Similarly if detector $D_1$ is a photographic plate the "no dot" would be a mirage photon accompanied by a nega-photon, as implicated in the above "negative absorption" experiment [14]. Indeed Quantum Oblivion [41], which underlies several quantum phenomena from the quantum Zeno to the Aharonov-Bohm effects, has shown, also with pre- and post-selections and "strong" measurements, how each such an apparent non-event can be decomposed into its occurrence followed by "un-occurrence" [21,22]. Nega-particles may thus become a common currency in quantum transactions. A profound time-symmetry of quantum reality seems to underlie the (in)famous asymmetry of measurement and classical reality.

## 8. Discussion: Time-Symmetric Causality and the Particle-Based Heisenberg Representation

Finally, it is instructive to point out that this picture of weak reality with the phenomena derived from it accords well with the Heisenberg approach advocated in [42-44]. We have pointed out that the Schrödinger wavefunction is often conceptually confusing and the Heisenberg operator-based formalism is more natural. According to this Heisenbergian view, a set of deterministic operators carries the same amount of information stored in the wavefunction, but in contrast to the latter can be viewed as a proper description of the single particle. Interestingly, these operators often obey nonlocal equations of motion [42-44], naturally accounting for quantum phenomena such as the Aharonov-Bohm effect, IFM and many others.

In the above examples too, it is the particle, with its host of mirage particles, rather than wave-like properties, which explain curious effects such as those manifested in the above experiments.

Several other derivations based on the TSVF, some already experimentally validated, have demonstrated an interesting "Cheshire cat" effect [18,45]: An apparently-intrinsic property of the particle, such as its spin, can traverse an MZI path *other* than that traversed by the particle itself. These phenomena can be more naturally understood in a particle-based framework: A massive particle traverses one arm of the MZI, while a pair of mirage-nega-mirgae, having zero mass but non-zero spin, traverses the other.

Deriving pairs of mirage particles and their accompanying nega-mirage particles, we can maintain an intuitive picture of continuous trajectories within the weak reality of quantum mechanics. Since the positive and negative mirage particles can hide one another, this picture allows us to think of the pre-selection event as the source for all the extra particles emerging from the original. Each then follows some definite trajectory through space-time until they all meet again at the post-selection where they are re-absorbed. The reabsorption is essential because this is where the back action from the pointer system on each of the different mirage particles collectively effects the original particle.

This picture has naturally emerged for the two-vector account of quantum processes, which is equivalent to the mainstream, one-vector account. It is worth comparing the two views.

*i)* Quantum measurement outcomes are not fully determined by the past. The future also takes part in shaping them. When the initial and final boundary conditions are an unlikely pair, Nature, so to speak, "goes out of its way" to reconcile between the forward and backward components of the resulting evolution, by giving rise to weak values like mirage and nega-mirage particles. Delicate measurements can later validate such phenomena that have occurred between pre- and post-selections.

*ii)* Among all (forward-in-time) possible quantum evolutions, there are some that involve anomalous weak values. These, however, are mixed with all the other weak values, stemming from all possible post-selections, and cannot be distinguished in real time (therefore giving rise to the customary expectation values). It is only the actual post-selection in a given experiment which informs us, in retrospect, which are the cases where these values were certainly involved.

Which account is more natural is a matter of personal choice. We only point out that it was the two-vector account which has revealed these phenomena based on weak values, the conventional alternative following only with hindsight.

To conclude, we have discussed a few thought experiments leading to a new perspective on the TSVF in particular, and on quantum mechanics in general. This particle-based approach is time-symmetric and realistic. Admittedly, the outlined picture, based on mirage and nega-mirage particles, is still far from being complete. Further derivations, experiments and generalizations are currently under work.

**Acknowledgments:** It is a pleasure to thank Andrei Khrennikov for organizing the series of conferences where these ideas were presented and for his support. We wish to thank Ryo Okamoto, Shigeki Takeuchi and Lev Vaidman for many helpful discussions. This research was funded by the Israel Science Foundation (Grant 1311/14), ICORE Excellence Center "Circle of Light," the German–Israeli Project Cooperation (DIP), the Faculty of Engineering in Bar-Ilan University, and the Fetzer Franklin Fund of the John E. Fetzer Memorial Trust.

**References**

1. Aharonov, Y.; Bergmann, P.G.; Lebowitz, J.L. Time symmetry in the quantum process of measurement. *Phys. Rev. B* **1964**, *134*, 1410–1416, DOI:10.1103/PhysRev.134.B1410.
2. Aharonov Y.; Vaidman, L. The two-state vector formalism of quantum mechanics: an updated review. In *Time in Quantum Mechanics*; Muga, J.G., Sala Mayato, R., Egusquiza. I.L., Eds.; Springer: Berlin/Heidelberg, Germany, 2002; pp. 369–412, ISBN: 978-3-540-43294-4.

3. Aharonov, Y.; Cohen, E.; Landsberger, T. The Two-Time Interpretation and Macroscopic Time-Reversibility. *Entropy* **2017**, *19*, 111, DOI:10.3390/e19030111
4. Aharonov, Y.; Albert, D.Z.; Vaidman, L. How the result of a measurement of a component of a spin 1/2 particle can turn out to be 100? *Phys. Rev. Lett.* **1988**, *60*, 1351–1354, DOI: 10.1103/PhysRevLett.60.1351.
5. Tamir, B.; Cohen, E. Introduction to weak measurements and weak values. *Quanta* **2013**, *2*, 7-17, DOI: 10.12743/quanta.v2i1.14.
6. Dressel, J. *et al.* Colloquium: Understanding quantum weak values: Basics and applications. *Rev. Mod. Phys.* **2014**, *86*, 307, DOI:10.1103/RevModPhys.86.307.
7. Vaidman, L. *et al.* Weak value beyond conditional expectation value of the pointer readings. *Phys. Rev. A* **2017**, *96*, 032114, DOI: 10.1103/PhysRevA.96.032114.
8. Aharonov, Y.; Vaidman, L. Complete description of a quantum system at a given time. *J. Phys. A* **1991**, *24*, 2315, DOI:10.1088/0305-4470/24/10/018.
9. Aharonov, Y.; Rohrlich, D. Quantum Paradoxes: Quantum Theory for the Perplexed; Wiley: New York, NY, USA, 2005, 978-3-527-40391-2.
10. Cohen, E.; Aharonov, Y. Quantum to classical transitions via weak measurements and post-selection. In *Quantam Structural Studies: Classical Emergence from the Quantum Level;* Kastner, R.E., Jenkic-Dugic, J., Jaroszkiewicz G., Eds.; Worlds Scientific, Singapore, 2017, pp. 401-425, ISBN: 978-1-78634-140-2.
11. Jozsa, R. Complex weak values in quantum measurement. *Phys. Rev. A* **2007,** *76*, 044103, DOI: 10.1103/PhysRevA.76.044103.
12. Berry M.V.; Shukla P. Typical weak and superweak values. *J. Phys. A* **2010**, *43*, 354024, DOI:10.1088/1751-8113/43/35/354024.
13. Hosoya, A.; Shikano, Y. Strange weak values. *J. Phys. A* **2010**, *43*, 385307. DOI: 10.1088/1751-8113/43/38/385307.
14. Aharonov, Y.; Cohen, E.; Carmi, A.; Elitzur, A.C. Extraordinary interactions between light and matter determined by anomalous weak values. *Proc. Roy. Soc. A.* **2018**, *474*, 20180030, DOI:10.1098/rspa.2018.0030.
15. Aharonov, Y.; Cohen, E.; Ben-Moshe, S. Unusual interactions of pre-and-post-selected particles. *EPJ Web Conf.* **2014**, *70, 00053*, DOI:10.1051/epjconf/20147000053.
16. Vaidman, L. Past of a quantum particle. *Phys. Rev. A* **2013**, *87*, 052104, DOI: 10.1103/PhysRevA.87.052104.
17. Danan, A.; Farfurnik, D.; Bar-Ad S.; Vaidman, L. Asking photons where have they been. *Phys. Rev. Lett.* **2013**, *111*, 240402, DOI:10.1103/PhysRevLett.111.240402.
18. Aharonov, Y.; Popescu, S.; Rohrlich, D.; Skrzypczyk, P. Quantum Cheshire cats. *New J. Phys.* **2013**, *15*, 113015, DOI: 10.1088/1367-2630/15/11/113015.
19. Aharonov, Y.; Cohen, E.; Landau, A.; Elitzur, A.C. The case of the disappearing (and re-appearing) particle. *Sci. Rep.* **2017**, *531*, DOI:10.1038/s41598-017-00274-w.
20. Elitzur, A.C.; Cohen, E.; Okamoto, R.; Takeuchi, S. Nonlocal position changes of a photon revealed by quantum routers. *Sci. Rep.* **2018**, *8*, 7730, DOI:10.1038/s41598-018-26018-y.
21. Cohen E.; Elitzur A.C. Voices of silence, novelties of noise: oblivion and hesitation as origins of quantum mysteries. *J. Phys. Conf. Ser.* **2015**, *626*, 012013, DOI:10.1088/1742-6596/626/1/012013.
22. Elitzur A.C.; Cohen E. 1 – 1 = Counterfactual: On the Potency and Significance of Quantum Non-Events. *Philos. Trans. Roy. Soc. A* **2016**, 374, *20150242*, DOI:10.1098/rsta.2015.0242.
23. Duprey, Q.; Matzkin, A. Null weak values and the past of a quantum particle. *Phys. Rev. A* **2017**, *95*, 032110, DOI:10.1103/PhysRevA.95.032110.
24. Matzkin, A. Weak values and quantum properties. arXiv:1808.09737.
25. Johansen, L.M.; Luis, A. Nonclassicality in weak measurements. *Phys. Rev. A* **2004**, 70, 052115, DOI: 10.1103/PhysRevA.70.052115.
26. Sokolovski, D. Weak values, "negative probability," and the uncertainty principle. *Phys. Rev. A* **2007,** *76*, 042125, DOI:10.1103/PhysRevA.76.042125
27. Sokolovski D. Are the weak measurements really measurements?. *Quanta* **2013**, 2, 50-57. DOI: 10.12743/quanta.v2i1.15.

28. Svensson, B.E. What is a quantum-mechanical "weak value" the value of?. *Found. Phys.* **2013**, *43*, 1193-1205, DOI:10.1007/s10701-013-9740-6.
29. Kastner R.E. Demystifying weak measurements. Found. Phys. **2017**, *47*, 697–707, DOI: 10.1007/s10701-017-0085-4.
30. Cohen, E. What weak measurements and weak values really mean: reply to Kastner. *Found. Phys.* **2017**, *47*, 1261-1266, DOI:10.1007/s10701-017-0107-2.
31. Li, F.; Hashmi F.A.; Zhang, J.X.; Zhu, S.Y. An ideal experiment to determine the 'past of a particle' in the nested Mach-Zehnder interferometer. *Chin. Phys. Lett.* **2015**, *32*, 050303, DOI: 10.1088/0256-307X/32/5/050303.
32. Ben-Israel, A. *et al.* An improved experiment to determine the 'past of a particle' in the nested Mach–Zehnder interferometer. *Chin. Phys. Lett.* **2017**, *34*, 020301, DOI: 10.1088/0256-307X/34/2/020301.
33. Englert, B.G.; Horia, K.; Dai, J.; Len, Y.L.; Ng, H.K. Past of a quantum particle revisited. *Phys. Rev. A* **2017**, *96*, 022126, DOI: 10.1103/PhysRevA.96.022126
34. Peleg, U.; Vaidman, L. Comment on " Past of a quantum particle revisited". arXiv:1805.12171.
35. de Laplace, P. S. *Théorie analytique des probabilités.* Courcier, Paris, France, 1812.
36. Okamoto, R.; Takeuchi, S. Experimental demonstration of a quantum shutter closing two slits simultaneously. *Sci. Rep.* **2016**, *6*, 35161, DOI:10.1038/srep35161.
37. Aharonov, Y.; Vaidman, L. How one shutter can close N slits. *Phys. Rev. A* **2003**, *67*, 042107, DOI: 10.1103/PhysRevA.67.042107.
38. Ball, P. Quantum Physics May Be Even Spookier Than You Think. Scientific American. Available online: https://www.scientificamerican.com/article/quantum-physics-may-be-even-spookier-than-you-think/.
39. Aharonov, Y.; Botero, A.; Nussinov, S.; Popescu, S.; Tollaksen, J.; Vaidman, L. The classical limit of quantum optics: not what it seems at first sight. *New J. Phys.* **2013**, *15*, 093006, DOI:10.1088/1367-2630/15/9/093006.
40. Elitzur, A. C.; Vaidman, L. Quantum mechanical interaction-free measurements. *Found. Phys.* **1993**, *23*, 987-997, DOI:10.1007/BF00736012.
41. Elitzur A.C.; Cohen E. Quantum Oblivion: A Master Key for Many Quantum Riddles. *Int. J. Quantum Inf.* **2015**, *12*, 1560024, DOI:10.1142/S0219749915600242.
42. Aharonov, Y.; Landsberger, T.; Cohen, E. A nonlocal ontology underlying the time-symmetric Heisenberg representation. arXiv:1510.03084.
43. Aharonov, Y. *et al.* Finally making sense of the double-slit experiment. *Proc. Natl. Acad. Sci.* **2017**, *114*, 6480-6485, DOI:10.1073/pnas.1704649114.
44. Aharonov Y.; Cohen E.; Elitzur A.C. Beyond Wavefunctions: A Time-Symmetric Nonlocal Ontology for Quantum Mechanics. In *Encouraging Openness - Essays for Joseph Agassi on the Occasion of His 90th Birthday*, Bar-Am, N., Gattei S. Eds., Springer 2017, 978-3-319-57668-8, pp. 235-239.
45. Denkmayr, T. *et al.* Observation of a quantum Cheshire Cat in a matter-wave interferometer experiment. *Nat. Commun.* **2014**, *5*, 5492, DOI:10.1038/ncomms5492.